\documentclass[english]{article}
\usepackage[T1]{fontenc}
\usepackage[latin9]{inputenc}
\usepackage{esint}

\makeatletter

\newcommand{\lyxaddress}[1]{
\par {\raggedright #1
\vspace{1.4em}
\noindent\par}
}

\usepackage{babel}
\makeatother

\begin{document}

\title{Neutrino Mixing and Cosmological Constant above GUT Scale}

\author{Bipin Singh Koranga }

\maketitle

\lyxaddress{Department of Physics, Kirori Mal college (University of Delhi,)
Delhi-110007, India}

\begin{abstract}
Neutrino mixing lead to a non zero contribution to the cosmological
constant. We consider non renormalization $1/M_{x}$ interaction term
as a perturbation of the neutrino mass matrix. We find that for the
degenerate neutrino mass spectrum. We assume that the neutrino masses
and mixing arise through physics at a scale intermediate between Planck
Scale and the electroweak scale. We also assume, above the electroweak
breaking scale, neutrino masses are nearly degenerate and their mixing
is bimaximal. Quantum gravitational (Planck scale )effects lead to
an effective $SU(2)_{L}\times U(1)$ invariant dimension-5 Lagrangian
involving neutrino and Higgs fields, which gives rise to additional
terms in neutrino mass matrix. There additional term can be considered
to be perturbation of the GUT scale bi-maximal neutrino mass matrix.
We assume that the gravitational interaction is flavour blind and
we study the neutrino mixing and cosmological constant due to physics
above the GUT scale.

\end{abstract}

\section{Introduction}

The problem of cosmological constant is currently one of the most
challenging open issue in theortical physics and cosmology. The main
difficulty comes from the mis match between theortical and accepted
number. Cosmology constant may arise from neutrino mixing {[}1]. In
this case of neutrinos, cosmological density reletated to the mixing
and mass difference among the different generations. Phenomenological
consequences of non-trival condensate structure of the flavour vacuum
have been studied for neutrino oscillations and Beta decay {[}2.3].
The nature of the cosmology constant $\Lambda$ is one of the most
intersting issues in modern theortical physics ans cosmology. Experimental
data coming from observation indicates that not only $\Lambda$ is
different from zero, $\Lambda$ also dominates the universe dynamics
driving an accelerated expansion {[}4,5]. In this paper, we study
the neutrino mixing due to Planck scale and contribution to cosmological
constant. In Section 2, we summarige the neutrino mixing due to Planck
scale effects. In Section 3, we discuss the neutrino mixing and cosmological
constant due to Planck scale effects. Section 4 is devoted to the
conclusions.

\section{Neutrino Oscillation Parameter due to Planck Scale Effects}

The neutrino mass matrix is assumed to be generated by the see saw
mechanism {[}6,7,8]. We assume that the dominant part of neutrino
mass matrix arise due to GUT scale operators and the lead to bi-maximal
mixing. The effective gravitational interaction of neutrino with Higgs
field can be expressed as $SU(2)_{L}\times U(1)$ invarinat dimension-5
operator {[}8],

\begin{equation}
L_{grav}=\frac{\lambda_{\alpha\beta}}{M_{pl}}(\psi_{A\alpha}\epsilon\psi_{C})C_{ab}^{-1}(\psi_{B\beta}\epsilon_{BD}\psi_{D})+h.c.\end{equation}

Here and every where we use Greek indices $\alpha,\,\beta$ for the
glavour states and Latin indices i,j,k for the mass states. In the
above equation $\psi_{\alpha}=(\nu_{\alpha},l_{\alpha})$is the lepton
doublet, $\phi=(\phi^{+},\phi^{o})$is the Higgs doublet and $M_{pl}=1.2\times10^{19}GeV$is
the Planck mass $\lambda$ is a $3\times3$ matrix in a flavour space
with each elements $O(1)$. The Lorentz indices $a,b=1,2,3,4$ are
contracted with the charge conjugation matrix $C$ and the $SU(2)_{L}$
isospin indices $A,B,C,D=1,2$ are contracted with $\epsilon=i\sigma_{2},\,\,\sigma_{m}(m=1,2,3)$are
the pauli matrices. After spontaneous electroweak symmetery breaking
the lagrangian in eq(1) generated additional term of neutrino mass
matrix

\begin{equation}
L_{mass}=\frac{v^{2}}{M_{pl}}\lambda_{\alpha\beta}\nu_{\alpha}C^{-1}\nu_{\beta},\end{equation}

where $v=174GeV$ is the $VEV$ of electroweak symmetric breaking.
We assume that the gravitational interaction is''flavour blind''
that is $\lambda_{\alpha\beta}$ is independant of $\alpha,\,\beta$indices.
Thus the Planck scale contibution to the neutrino mass matrix is

\begin{equation}
\mu\lambda=\mu\left(\begin{array}{ccc}
1 & 1 & 1\\
1 & 1 & 1\\
1 & 1 & 1\end{array}\right),\end{equation}

where the scale $\mu$ is 

\begin{equation}
\mu=\frac{v^{2}}{M_{pl}}=2.5\times10^{-6}eV.\end{equation}

We take eq(3) as perturbation to the main part of the neutrino mass
matrix, that is generated by GUT dynamics. To calculate the effects
of perturbation on neutrino observables. The calculation developed
in an earlier paper {[}8]. A natural assumption is that unperturbed
($0^{th}$ order mass matrix) $M$~is given by

\begin{equation}
\mathbf{M}=U^{*}diag(M_{i})U^{\dagger},\end{equation}

where, $U_{\alpha i}$ is the usual mixing matrix and $M_{i}$ , the
neutrino masses is generated by Grand unified theory. Most of the
parameter related to neutrino oscillation are known, the major expectation
is given by the mixing elements $U_{e3}.$ We adopt the usual parametrization.

\begin{equation}
\frac{|U_{e2}|}{|U_{e1}|}=tan\theta_{12},\end{equation}

\begin{equation}
\frac{|U_{\mu3}|}{|U_{\tau3}|}=tan\theta_{23},\end{equation}

\begin{equation}
|U_{e3}|=sin\theta_{13}.\end{equation}

In term of the above mixing angles, the mixing matrix is

\begin{equation}
U=diag(e^{if1},e^{if2},e^{if3})R(\theta_{23})\Delta R(\theta_{13})\Delta^{*}R(\theta_{12})diag(e^{ia1},e^{ia2},1).\end{equation}

The matrix $\Delta=diag(e^{\frac{1\delta}{2}},1,e^{\frac{-i\delta}{2}}$)
contains the Dirac phase. This leads to CP violation in neutrino oscillation
$a1$ and $a2$ are the so called Majoring phase, which effects the
neutrino less double beta decay. $f1,$ $f2$ and $f3$ are usually
absorbed as a part of the definition of the charge lepton field. Planck
scale effects will add other contribution to the mass matrix that
gives the new mixing matrix can be written as {[}8]

\[
U^{'}=U(1+i\delta\theta),\]

\[
\left(\begin{array}{ccc}
U_{e1} & U_{e2} & U_{e3}\\
U_{\mu1} & U_{\mu2} & U_{\mu3}\\
U_{\tau1} & U_{\tau2} & U_{\tau3}\end{array}\right)\]

\begin{equation}
+i\left(\begin{array}{ccc}
U_{e2}\delta\theta_{12}^{*}+U_{e3}\delta\theta_{23,}^{*} & U_{e1}\delta\theta_{12}+U_{e3}\delta\theta_{23}^{*}, & U_{e1}\delta\theta_{13}+U_{e3}\delta\theta_{23}^{*}\\
U_{\mu2}\delta\theta_{12}^{*}+U_{\mu3}\delta\theta_{23,}^{*} & U_{\mu1}\delta\theta_{12}+U_{\mu3}\delta\theta_{23}^{*}, & U_{\mu1}\delta\theta_{13}+U_{\mu3}\delta\theta_{23}^{*}\\
U_{\tau2}\delta\theta_{12}^{*}+U_{\tau3}\delta\theta_{23}^{*}, & U_{\tau1}\delta\theta_{12}+U_{\tau3}\delta\theta_{23}^{*}, & U_{\tau1}\delta\theta_{13}+U_{\tau3}\delta\theta_{23}^{*}\end{array}\right).\end{equation}

Where $\delta\theta$ is a hermition matrix that is first order in
$\mu${[}8,9]. The first order mass square difference $\Delta M_{ij}^{2}=M_{i}^{2}-M_{j}^{2},$get
modified {[}8,9] as

\begin{equation}
\Delta M_{ij}^{'^{2}}=\Delta M_{ij}^{2}+2(M_{i}Re(m_{ii})-M_{j}Re(m_{jj}),\end{equation}

where

\[
m=\mu U^{t}\lambda U,\]

\[
\mu=\frac{v^{2}}{M_{pl}}=2.5\times10^{-6}eV.\]

The change in the elements of the mixing matrix, which we parametrized
by $\delta\theta${[}8,9], is given by

\begin{equation}
\delta\theta_{ij}=\frac{iRe(m_{jj})(M_{i}+M_{j})-Im(m_{jj})(M_{i}-M_{j})}{\Delta M_{ij}^{'^{2}}}.\end{equation}

The above equation determine only the off diagonal elements of matrix
$\delta\theta_{ij}$. The diagonal element of $\delta\theta_{ij}$
can be set to zero by phase invariance. Using Eq(10), we can calculate
neutrino mixing angle due to Planck scale effects,

\begin{equation}
\frac{|U_{e2}^{'}|}{|U_{e1}^{'}|}=tan\theta_{12}^{'},\end{equation}

\begin{equation}
\frac{|U_{\mu3}^{'}|}{|U_{\tau3}^{'}|}=tan\theta_{23}^{'},\end{equation}

\begin{equation}
|U_{e3}^{'}|=sin\theta._{13}^{'}\end{equation}

For degenerate neutrinos, $M_{3}-M_{1}\cong M_{3}-M_{2}\gg M_{2}-M_{1},$
because $\Delta_{31}\cong\Delta_{32}\gg\Delta_{21}.$ Thus, from the
above set of equations, we see that $U_{e1}^{'}$ and $U_{e2}^{'}$
are much larger than $U_{e3}^{'},\,\, U_{\mu3}^{'}$ and $U_{\tau3}^{'}$.
Hence we can expect much larger change in $\theta_{12}$ compared
to $\theta_{13}$ and $\theta_{23}\,[10].$ As one can see from the
above expression of mixing angle due to Planck scale effects, depends
on new contribution of mixing $U^{'}=U(1+i\delta\theta).$

\section{Neutrino Mixing and Cosmological Constant Due to Planck Scale Effects}

The connection between the vacuum energy density $<\rho_{vac}>$and
the cosmology constant $\Lambda$ is provided by the well known relation

\begin{equation}
<\rho_{vac}>=\frac{\Lambda}{4\pi G},\end{equation}

where $G$ is the gravitational constant.

The expression of vacuum energy density $<\rho_{vac}^{mix}>$ due
to neutrino mixing is given by {[}11,12,13]

\begin{equation}
<\rho_{vac}^{mix}>=32\pi^{2}sin^{2}\theta_{12}\int dkK^{2}(\omega_{k,1}+\omega_{k,2})|V_{k}|^{2},\end{equation}

If we chose $K\gg\sqrt{m_{1}m_{2}}$ , we obtain

\begin{equation}
<\rho_{vac}^{mix}>=\propto sin^{2}\theta_{12}(m_{2}-m_{1})^{2}K^{2}\intop_{0}^{k}dkk^{2}(\omega_{k,1}+\omega_{k,2})|V_{K}|^{2},\end{equation}

For hierarchical neutrino model, for which $m_{2}>m_{1}$, wehave
in this case $K\gg\sqrt{m_{1}m_{2}}$ and take into account the asymptotic
properties of $V_{k}$

\[
|V_{k}|^{2}\simeq\frac{(m_{2}-m_{1})^{2}}{4K^{2}},\,\,\,\,\, K\gg\sqrt{m_{1}m_{2}}\]
.

We get

\begin{equation}
<\rho_{vac}^{mix}>=\propto sin^{2}\theta_{12}(m_{2}^{2}-m_{1}^{2})\propto\frac{\Lambda}{4\pi G},\end{equation}

The new cosmological constant $\Lambda$ due to Planck scale effects
is given by

\begin{equation}
\Lambda^{'}=\propto sin^{2}\theta_{12}^{'}(m_{2}^{2}-m_{1}^{2}),\end{equation}

where $\theta_{12}^{'}$ is given by eq(13)

We consdier the Planck scale effects on neutrino mixing and we get
the given range of mixing parameter of MNS matrix

\begin{equation}
U^{'}=R(\theta_{23}+\epsilon_{3})U_{phase}(\delta)R(\theta_{13}+\epsilon_{2})R(\theta_{12}+\epsilon_{1}).\end{equation}

In Planck scale, only $\theta_{12}$($\epsilon_{1}=\pm3^{o})$have
resonable deviation and $\theta_{23},\,\theta_{13}$ deviation is
very small less than $0.3^{o}${[}10]. In the new mixing at Planck
scale we get he cosmological density

\begin{equation}
\Lambda^{'}=\propto sin^{2}(\theta_{12}^{}\pm\epsilon_{1})(m_{2}^{'2}-m_{^{'}1}^{2}),\end{equation}

The presence of a cosmological constant fluid has to be comptible
with the structure formation, allow to set the upper bound $\Lambda<10^{-56}cm^{-2}[14]$.
Due to Planck scale effects mixing angle $\theta_{12}$ deviated the
cosmological constant $\Lambda.$

\section{Conclusions}

We assume that the main part of neutrino masses and mixing from GUT
scale operator. We considered these to be $0^{th}$ order quanties.
We further assume that GUT scale symmetery constrain the neutrino
mixing angles to be bimaximal. THe gravitational interaction of lepton
field with S.M Higgs field give rise to a $SU(2)_{L}\times U(1)$
invirant dimension-5 effective lagragian give originally by Wenberg
{[}15]. On electroweak symmetery breaking this operators leads to
additional mass terms. We considred these to be perturbation of GUT
scale mass terms. We compute the first order correction to neutrino
mass eigen value and mixing angles. In {[}10], it was shown that the
change in $\theta_{13},\,\theta_{23}$ is very small (less then $0.3^{o})$but
the change in $\theta_{12}$ can be substantial about $\pm3^{o}.$THe
change in all the mixing angle are proportinal to the neutrino mass
eigenvalues. To maximizer the change, we assumed degenerate neutrino
mass $2.0eV$. For degenerate neutrino masses, the change in $\theta_{13},\,\theta_{23}$
are inversely propertional to $\Delta_{21}.$Since $\Delta_{31}\cong\Delta_{32}\gg\Delta_{21}.$
the change in $\theta_{12}$is much larger than the change in other
mixing angle. In this paper, we write the cosmological constant above
GUT scale in term of mixing angle for Majarona neutrinos, these expression
in eq(x) for vacuum mixing. For Majarona neutrino, the expression
is $\Lambda=\propto sin^{2}\theta_{12}^{'}(m_{2}^{2}-m_{1}^{2}),$.
In this paper, finially we wish make a important comment. Due to Planck
scale effects mixing angle $\theta_{12}$deviated the cosmological
constant $\Lambda$.


\begin{thebibliography}{10}
\bibitem{key-8}M. Blasone, $et\, al.,$Phys.Lett.\textbf{A}323,182(2004). 

\bibitem{key-9}M. Balsone, $et\, al.,$ Phys. Rev.\textbf{D}.67,073011
(2003).

\bibitem{key-1}M. Blasone, $et\, al.,$hep-ph/0307205..

\bibitem{key-3}S. Perlmutter $et\, al.,$ Astrophys.\textbf{J}.517,565
(1999).

\bibitem{key-10}V. Sahni and A. Starobinsky, Int.J.Mod.Phys.\textbf{D9},373
(2000).

\bibitem{key-1}R.N Mohapatra $et\, al.,$Phys.Rev.Lett \textbf{44},912
(1980).

\bibitem{key-10}S. Coleman and S. L Galshow, Phys. Rev \textbf{D}
59,116008 (1999).

\bibitem{key-11}F. Vissani $et\, al.,$Phys.Lett. B571, 209, (2003).

\bibitem{key-12}Bipin Singh Koranga, Mohan Narayan and S. Uma Sankar,
arXiv:hep-ph/0611186.

\bibitem{key-14}Bipin Singh Koranga, Mohan Narayan and S. Uma Sankar,Phys.Lett.\textbf{B665},
63 (2008).

\bibitem{key-4}M. Blasone, $et\, al.,$Phys.Lett. \textbf{A}323,
182 (2004).

\bibitem{key-5}M. Blasone, $et\, al.,$Braz.\textbf{J}.Phys.35:447-454
(2005).

\bibitem{key-6}A.Capolupo, S.Capozziello and G.Vitiello, Phys. Lett.
\textbf{A}363,53 ( 2007).

\bibitem{key-13}Ya.B. Zeldovich, I.D. Novikov, Structure and evolution
of the universe Moscow, Izdatelstvo Nauka (1975).

\bibitem{key-3}S. Weinberg, Phys.Rev.Lett.\textbf{43}.1566 (1979).

\end{thebibliography}
\end{document}